\documentclass[sigconf]{acmart}

\def\BibTeX{{\rm B\kern-.05em{\sc i\kern-.025em b}\kern-.08emT\kern-.1667em\lower.7ex\hbox{E}\kern-.125emX}}
\usepackage{hyperref}
\usepackage{caption}
\usepackage{subcaption}    
\usepackage{caption} 
\usepackage{times}
\usepackage{graphicx}
\usepackage{verbatim}
\usepackage{subcaption}
\usepackage{float}
\copyrightyear{2019}
\acmYear{2019}
\setcopyright{acmcopyright}
\acmConference[6th NSysS 2019]{6th International Conference on Networking, Systems and Security}{December 17--19, 2019}{Dhaka, Bangladesh}
\acmBooktitle{6th International Conference on Networking, Systems and Security (6th NSysS 2019), December 17--19, 2019, Dhaka, Bangladesh}
\acmPrice{15.00}
\acmDOI{10.1145/3362966.3362985}
\acmISBN{978-1-4503-7699-0/19/12}

\begin{document}

\title{As You Are, So Shall You Move Your Head: A System-Level Analysis between Head Movements and Corresponding Traits and Emotions}

\author{Sharmin Akther Purabi, Rayhan Rashed, Md. Mirajul Islam, Md. Nahiyan Uddin, Mahmuda Naznin, and A. B. M. Alim Al Islam}
\affiliation{%
  \institution{Department of Computer Science and Engineering, BUET, West Polashi, Dhaka, Bangladesh-1000}
  \country{Email: \{1505067.sap, 1505006.mrr, 1405119.mi, 1405102.mnu\}@ugrad.cse.buet.ac.bd,  \{mahmudanaznin, alim\_razi\}@cse.buet.ac.bd}
}

\renewcommand{\shortauthors}{Sharmin and Rayhan et al.}
\renewcommand{\shorttitle}{As You Are, So Shall You Move Your Head}
\newcommand*{\origrightarrow}{}
\let\oldarrow\textrightarrow
\renewcommand*{\textrightarrow}{\fontfamily{cmr}\selectfont\origrightarrow}

\begin{abstract}
%mirajul (revised)
%rayhan
Identifying physical traits and emotions based on system-sensed physical activities is a challenging problem in the realm of human-computer interaction. Our work contributes in this context by investigating an underlying connection between head movements and corresponding traits and emotions. To do so, we utilize a head movement measuring device called eSense, which gives acceleration and rotation of a head. Here, first, we conduct a thorough study over head movement data collected from $46$ persons using eSense while inducing five different emotional states over them in isolation. Our analysis reveals several new head movement based findings, which in turn, leads us to a novel unified solution for identifying different human traits and emotions through exploiting machine learning techniques over head movement data. Our analysis confirms that the proposed solution can result in high accuracy over the collected data. Accordingly, we develop an integrated unified solution for real-time emotion and trait identification using head movement data leveraging outcomes of our analysis.
\end{abstract}




\maketitle

\section{Introduction}
%mirajul (revised)
%rayhan
Identifying human traits and emotional states has numerous applications in diversified sectors covering health diagnosis, human-robot interactions, humanoid research, etc. Reasons behind the diversified applications is the fact that activity and behaviour vary from person to person, and the activity and behavior mostly get dictated by the person's trait and emotional state. Thus, it is imperative that if the traits and emotional state can get identified, corresponding activity and behavior can get predicted and generated. However, the task of identifying human traits and emotional states is always regarded as a challenging task, and perhaps this task is still at its embryonic stage exhibiting a substantial scope of conducting research.

\textit{Traits} carry most of the reasons behind diversified human behaviors and physical activities. Different methodologies have been explored to identify the relationship between the human activities and their connection with traits and emotional states \cite{joseph2016, guo2008}. Some of the human traits can be identified by some physical diagnosis processes such as blood test \cite{gnann2009}, dope test \cite{Connor871}, etc. Examples of such identifiable traits include smoking, taking alcohol, etc. Although these diagnosis processes are well accepted, they are often expensive to some extents \cite{ash2001}. On the other hand, questionnaire or interview-based approaches used  to identify the traits \cite{muba2011} can be considered to be cheaper alternatives. However, these approaches are trait-specific \cite{rog2005}, and thus, these lack potency to be scaled up. Accordingly, it is very difficult (if not impossible) to apply an interview-based approach to another application, as this demands to have new domain experts to   identify user traits and emotional states. Moreover, success of these approaches often depends on overcoming socio-cultural and other human related barriers, which often limits their applicability. Therefore, the questionnaire or interview-based approaches cannot be applied in a ubiquitous manner.

To this extent, we endeavour to explore devising a ubiquitous method of identifying human traits and emotional states going beyond the traditional tests and interview-based approaches. Here, we adopt a common human activity - head movement - to explore whether it gets related to human traits and emotional states. In case the head movement activity can get related to human traits and emotional states, we can attempt to identify the human traits and emotional states through analyzing corresponding head movement data. To the best of our knowledge, such an exploration is yet to be focused in the literature, and we are the first to do it.

Accordingly, in this paper, we conduct a research on revealing relationships between human traits and head movement data through analyzing \textit{translational} and \textit{rotational} human-head movement data. We apply machine learning techniques to relate human traits with the corresponding head movement data. Here, we build our data set by collecting head movement data from $46$ participants using a device called \textit{eSense}. The device has built-in Accelerometer and Gyroscope in it \cite{kawsar2018esense}. We collect data with the device while imposing five different emotional states in isolation on each of the participants. Subsequently, we feed different machine learning models with our collected data set to analyze underlying relationships between the head movement data and corresponding traits and emotional states.
Our study reveals that there exist relationships between spontaneous movement of the head and corresponding traits and emotional states.

Our collected head movement data consists of \textit{translational} and \textit{rotational} motion along the three dimensional physical axes. 

Here, we explore $14$ different human traits. Examples of the traits are religion practice, religious belief, religion practice among family members, physical exercise, smoking, diabetes among family members, heart disease among family members, brain stroke among family members,fastfood intake habits etc. Besides, we explore five different emotional states such as happy, sad, etc.
We find out relationships between the head movement data and the corresponding traits and emotional states through utilizing machine learning techniques.
Our machine learning-based analysis over the collected data demonstrates substantial accuracy in identifying the traits and emotional states.

Based on our work in this paper, we make the following set of contributions:

\begin{itemize}
\item
We collect head movement data (at the granularity of each second) with eSense device. While collecting the data, we induce five different emotional states in isolation over each of our $46$ participants by showing related videos.
\item 
To expedite the process of data collection, we develop an automated data collection application to eventually come up with an automated integrated survey system.
\item 
We apply machine learning based techniques over the collected data to analyze how the head movement data get correlated with the traits and emotional states. Here, we analyze capability of the machine learning techniques in digging the correlation, and thus performing identification of traits and emotional states based on head movement data.
\end{itemize}

The organization of this paper is as follows. In Section 2, we discuss relevant research studies. In Section 3, we provide the detail methodology of our approach. Later, in Section 4, we present our experimental results. Finally, in Section 5, we conclude this paper with future research directions.

\section{Background Study}
%mirajul
Our research builds on prior examinations of head movement to indicate it's relationship with human emotions and other physical activities. The study by Wallbot et al. demonstrated that body movements and postures to some degree are specific for certain emotions\cite{wallbott1998}. A sample of 224 video takes, in which actors and actresses portrayed the emotions of elated joy, happiness, sadness, despair, fear, terror, cold anger, hot anger, disgust, contempt, shame, guilt, pride, and boredom via a scenario approach, was analyzed using coding schemata for the analysis of body movements and postures. Results indicate that some emotion-specific movement and posture characteristics seem to exist, but that for body movements differences between emotions can be partly explained by the dimension of activation. While encoder (actor) differences are rather pronounced with respect to specific movement and posture habits, these differences are largely independent from the emotion-specific differences found. The results are discussed with respect to emotion-specific discrete expression models in contrast to dimensional models of emotion encoding. When a person is stuck with grief, normally he/she shows lethargy which leads them to make less human body movement. On the other hand, being happy one normally talks with excitement and a jolly mind. Then their physical body movement increases. Our motivation is to explore whether similar conclusions can be drawn for head movement patterns also. For this purpose we intend to go through a rigorous research methodology. Provided we get significant success, we can move on further to find co-relations between physical traits, emotions and head movement patterns. Throughout this research, our motivation was to check whether any co-relation exists and to what extent.

Many human physiological studies have provided strong evidence for a close link among human emotion, trait and head movement. The study in \cite{gunes2010} focused on dimensional prediction of emotions from spontaneous conversational head gestures. Their preliminary experiments show that it is possible to automatically predict emotions in terms of these five dimensions (arousal, expectation, intensity, power and valence) from conversational head gestures. A basic method of using temporal and dynamic design elements, in particular physical movements, to improve the emotional value of products is presented in \cite{lee2007}. To utilize physical movements in design, a relational framework between movement and emotion was developed as the first step of that research. In the framework, the movement representing emotion has been defined in terms of three properties: velocity, smoothness and openness. Based on this framework, a new interactive device, 'Emotion Palpus', was developed, and a user study was also conducted. The result of the research showed that emotional user experience got improved when used as a design method or directly applied to design practice as an interactive element of products.

In \cite{tanya2005}, the authors  examined whether it is possible to identify the emotional content of behaviour from point-light displays where pairs of actors are engaged in interpersonal communication. In this study \cite{munhall2004}, the authors suggested that nonverbal gestures such as head movements play a more direct role in the perception of speech than previously known. According to their study, the head movement correlated strongly with the pitch (fundamental frequency) and amplitude of the talker's voice. Sogon et al. \cite{sogon1989} showed that seven fundamental emotions of joy, surprise, fear, sadness, disgust, anger, contempt and three affective-cognitive structures for the emotions of affection, anticipation, and acceptance were displayed by four Japanese actors/actresses with their backs turned toward the viewer. The emotions of sadness, fear, and anger as expressed in kinetic movement showed high agreement between the two cultural groups. Joy and surprise, even though they are classified as fundamental emotions, contained some cultural components that affected the judgments. Furthermore, the subjects from USA successfully identified disgust as portrayed by Japanese actors/actresses, but the subjects from Japan did not identify the expressions of disgust or contempt. Affection, anticipation, and acceptance have some cultural components that are interpreted differently by Japanese and Americans, and this accounted for some of the misunderstandings. Nevertheless, most of the scenes depicting emotions and affective-cognitive structures emotions were correctly identified by the subjects of each culture.
There is another research work \cite{jari2008}, which compared the identification of basic emotions from both natural and synthetic dynamic vs. static facial expressions in 54 subjects. They found no significant differences in the identification of static and dynamic expressions from natural faces. In contrast, some synthetic dynamic expressions were identified much more accurately than static ones. This effect was evident only with synthetic facial expressions whose static displays were non-distinctive. Their results showed that dynamics does not improve the identification of already distinctive static facial displays. On the other hand, dynamics had an important role for identifying subtle emotional expressions, particularly from computer-animated synthetic characters.

%mirajul(revised)
%rayhan
\section{Our Proposed Approach}
Our working methodology can be divided in two phases: data collection and training phase, testing phase. We describe the phases in detail in this section.

\subsection{Preliminaries}
Our proposed technique relies on training data, which is collected through the \textit{eSense} device which is a bluetooth enabled device. Collected data samples are stored and exploited to train a model. Training phase includes two steps: collecting training data and generating training results, selection of the model. The training phase begins with the respondent wearing the eSense device. The data is collected through its BLE interface and sensors and stored in an android device. Later on, the data is synchronized with the experimental timeline and processed for analyzing.

\begin{figure*}[!t]
    \centering
    \includegraphics[scale=0.40]{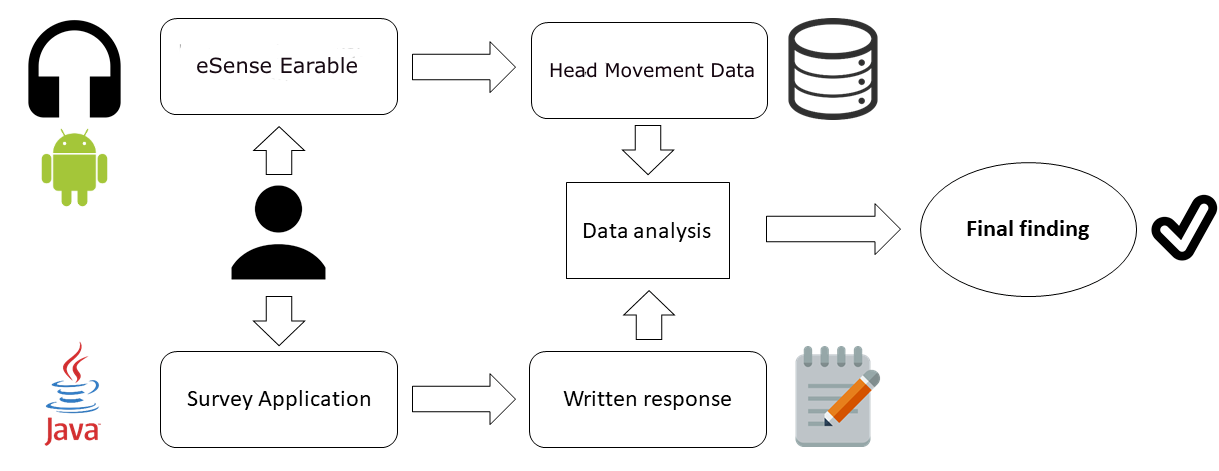}
    \caption{Workflow of our proposed system}
    \label{fig:flow}
\end{figure*}

\textit{Data Collection Phase}: From the earable, using its accelerometer, we get the translational motion component along the x, y, and z axes. Moreover, with built-in gyroscope, the rotational components of head movement along x, y, and z axes have been collected. We have computed  the mean and standard deviation of all data points. From the questionnaire responses, we have found demographic information of the participants and also their 14 human traits ~\cite{demo2009}. Individually no movement component show any significant outcome, but a combination of these components contribute to the identification of traits and emotion.
 
\textit{Training Phase}: Each head movement data point can be considered as a collection of 6 components, contributing to translational and rotational motion in the direction of x, y, and z axis. So, it can be considered as a 6 dimensional vector. Taking a combined translational and rotational term ($\sqrt{x^2+y^2+z^2}$) reduces the dimension of vector to 2. This transforms overall head movement data in translational and rotational form. We have tested with the later one\'s mean and standard deviation.

\textit{Testing Accuracy of the Trained Models}: After the training phase is done, it is time to check its consistency, correctness, and pick the best model that best describes the trait. Therefore, a sample data point goes through selected machine learning model to extract the feature. The original feature with the result obtained are compared.

\subsection{Data Collection}
In this section we describe the details of data collection.
\iffalse
%repetition
As it is needed to collect multiple sensor data simultaneously, a platform is necessary that can integrate both sensors' data and carefully run the survey system at a time. Moreover, storage for the collected data is also a priority. For this experiment, we used twofold hardware and applications. One Java application installed a computer to run the survey system, and an android app to collect the sensor data from the BLE interface is used together. Later these data were merged based on time-stamp values of events. The survey system consists of videos based on different human emotions. For analysis, we used the Auto-WEKA software package from WEKA-3-8-2. It finds out the best classifier or regression algorithm based on the training dataset. The more time Auto-WEKA runs, the better it gets with the choice classifier algorithm selection from a large pool of candidate algorithms. Our run time  for training was 30 minutes per trait per emotional state.
\fi

\subsubsection{Devices Used for Data Collection}
Each of the participants wore an earable called \textit{eSense}. This device is equipped with a built in sensors like accelerometer and gyroscope \cite{kawsar2018esense}. Data is transmitted from these sensors to an android device in real-time using Bluetooth Low Energy (BLE) radio connection. 
% DO WE REALLY NEED DEVICE'S Picture eSense Device?
\begin{figure}[!t]
  \centering
  \includegraphics[scale=0.25]{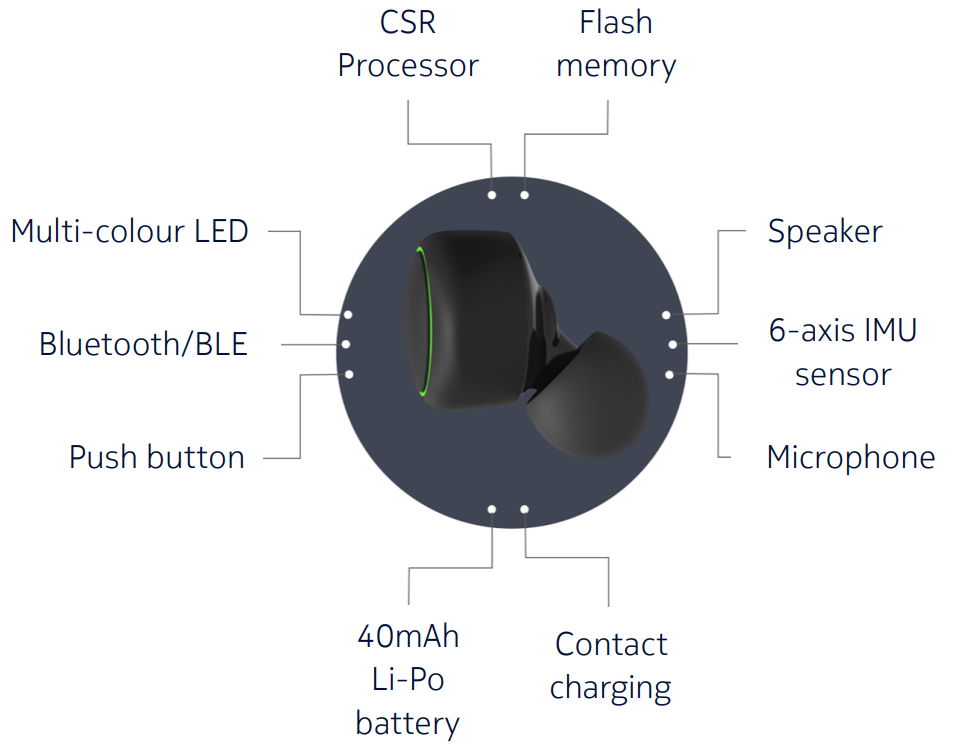}
  \caption{eSense device}
\end{figure}

\begin{figure}[htbp]
  \centering
  \includegraphics[scale=0.15]{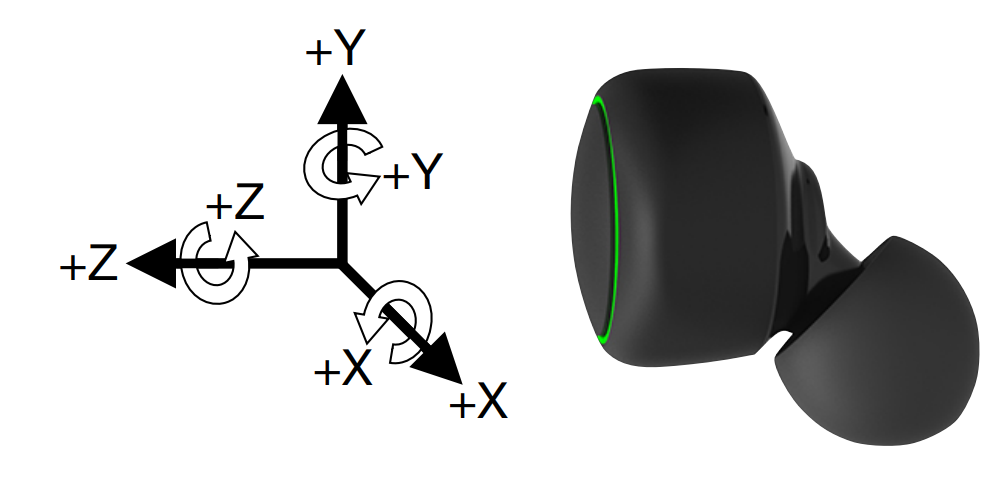}
  \caption{Axis orientation of eSense \cite{8490189}}
\end{figure}
We connect the device thorough using bluetooth to an Android 9 OS based smartphone.  Each participant sat before a computer which was used to communicate with and receive data from the sensor suites
of the earable. At the time of collecting brain wave data, five different videos were shown on the computer screen to create different emotional environments for the participant.
Once the brain data collection phase is finished, each participant filled up our questionnaire which provided few subjective data about him/her.

\begin{figure}[!t]
  \centering
  \includegraphics[scale=0.25]{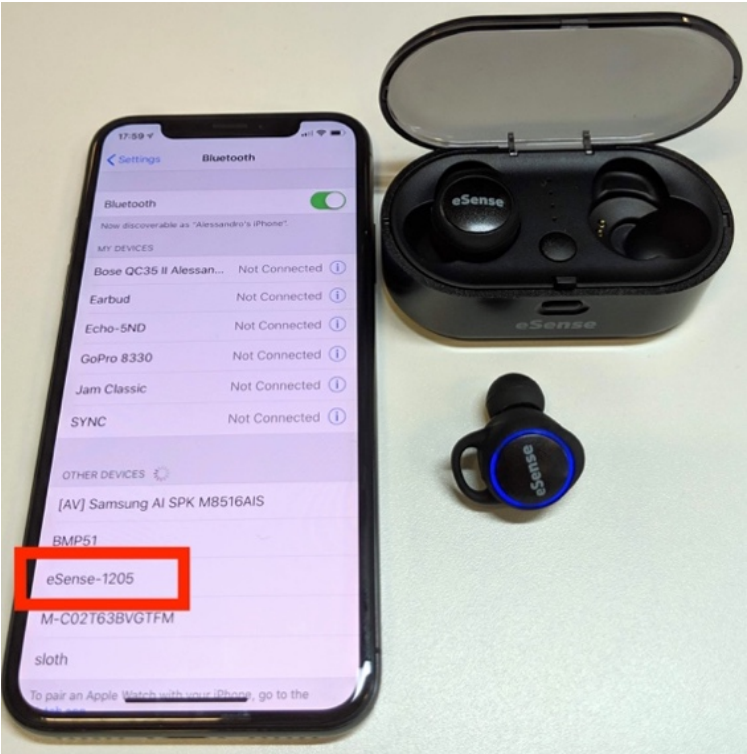}
  \caption{Pairing of eSense device}
\end{figure}

%[We need one more figure here]

\subsubsection{Customized Application for Data Collection}
Two different pool of applications were used for the whole process. After the initial bluetooth connection setup, an app was run to initiate the actual data collection. At the same time a java based desktop application was run to show the videos and run the survey.
Time-stamps were recorded for each data sampling time from android based device data and each phase of desktop application. Later collected datasets were integrated to the training phase.
\begin{figure}[!t]
     \centering
     \begin{subfigure}[b]{0.35\textwidth}
         \centering
         \includegraphics[width=\textwidth]{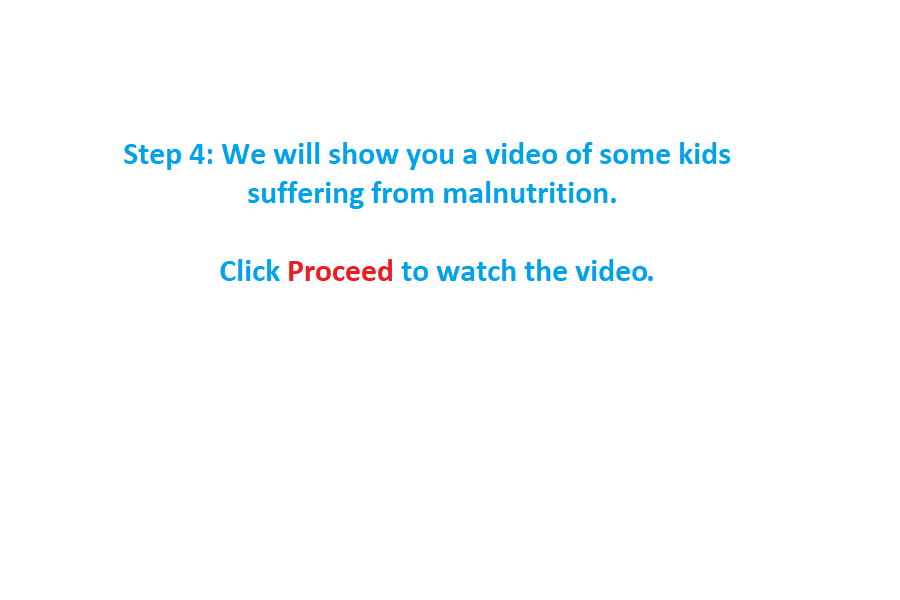}
         \caption{Notification for sad segment}
         \label{Sad Segment}
     \end{subfigure}
     \hfill
          \begin{subfigure}[b]{0.35\textwidth}
         \centering
         \includegraphics[width=\textwidth]{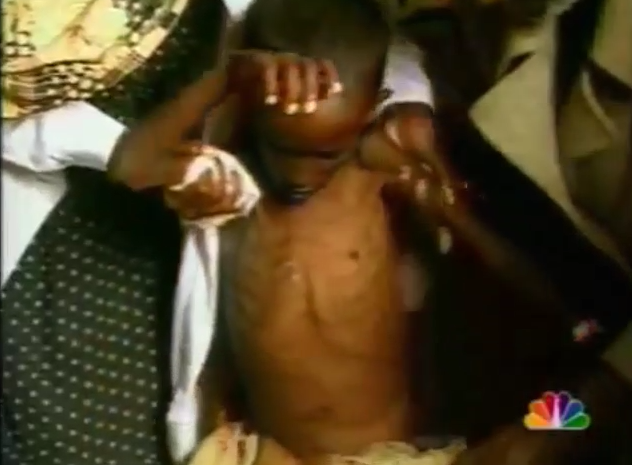}
         \caption{Sad Video}
         \label{Sad Video}
     \end{subfigure}
        \caption{Snapshots of desktop application}
        \label{Snapshot of Our Application}
\end{figure}

%mirajul(revised)
\subsubsection{Demography of the participants}
Participants for this experiment were selected from a diversified background. They cover different age groups (from 18 to 45), different genders (45\% female and 55\% male) and different levels of experience in behavioral perspective and computer literacy. 70\% of the respondents are frequent computer users, and the rest use computer occasionally. We tried to collect data to make a balanced data set for each trait, and we have successfully done it for all the 7 traits.

%mirajul(revised)
\subsubsection{Challenges in Data Collection} 
The data collection process needs a place where no external distraction is possible. Hands on head movement data extraction based on induced emotion is only possible if there is no other external factor, distraction responsible for slightest head movement. This makes the choice of setup environment very sophisticated. Data collection was only possible in controlled laboratory experiment.  

%mirajul(revised)
%purabi
\subsection{Machine Learning Based Analysis}
We used the Auto-Weka package of the Weka software 3.8.2 version. It finds out the best classifying or regression algorithm automatically given the training dataset. We run the Auto-Weka with two parallel core and approximately 30-40 minutes time was given to check sufficiently large number of combinations of algorithms.

As we collect head movement data showing five different videos, we get five different sets of accelerometer and gyroscope data and each of them is stored during a particular emotional state. We repeat our machine learning analysis with the five set of different emotional states' head movement data. From the survey questionnaire answered by the train participants, we chose 7 traits that we are looking for to train. We used participant's answer as the ground truth for all the trait and behavior related questions of the survey. For each trait we train one machine learning model where the label is the trait, and features are the mean and standard deviation of accelerometer and gyroscope data of participants as described before. As a result, we trained 7 machine learning models corresponding to 7 traits in each of the five emotional states which produces 35 training models. Among these 7 traits Smoking, Practicing religion, High-fat food intake habit, high-sugar food intake habit, heart diseases and diabetes in family are categorised in yes or no. But the fast-food intake data was categorised into 3 ranges (Low, Medium, High) based on how many fast-food meal participant take in each week on average. 

When we predict a certain trait during the evaluation period, we use an aggregated form of all the 5 emotional states\' models for the trait where we take the model with best training accuracy. We have also made a model to predict participant\'s emotion from his accelerometer and gyroscope data. We label the emotional state of a participant's data based on which video the participant was watching during that time. We chose a funny video of Mr.Bean for representing the \textit{happy} emotional state, a food preparing video with background sound for representing \textit{surprise} emotional state, a video (with command to keep close their eyes) with no sound and visual scenery for representing \textit{neutral state}, a video of people suffering from malnutrition for representing \textit{sad} emotional state and finally an animation video showing effects of smoking to human body for representing \textit{disgust}.

%mirajul(revised)
%purabi
\section{Experimental Results}
%\subsection
This section describes the findings of our experiments. 

\begin{figure}[H]
\centering
\subcaptionbox{Accelerometer data}{\includegraphics[width=0.23\textwidth]{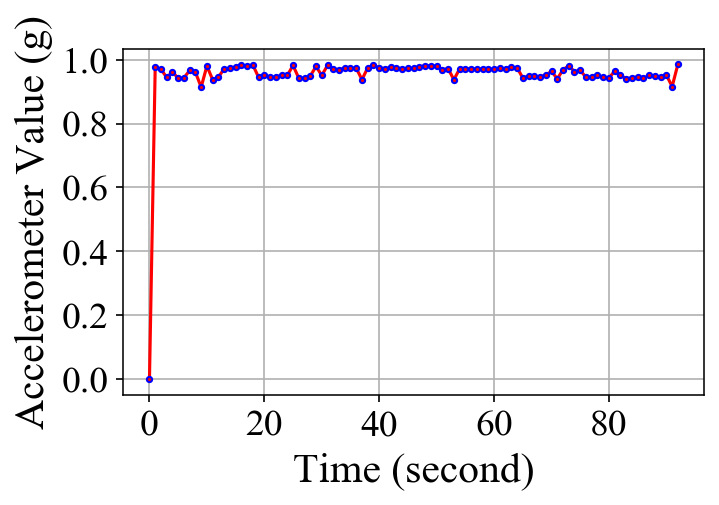}}%
\hfill % <-- Line break
\subcaptionbox{Gyroscope data}{\includegraphics[width=0.23\textwidth]{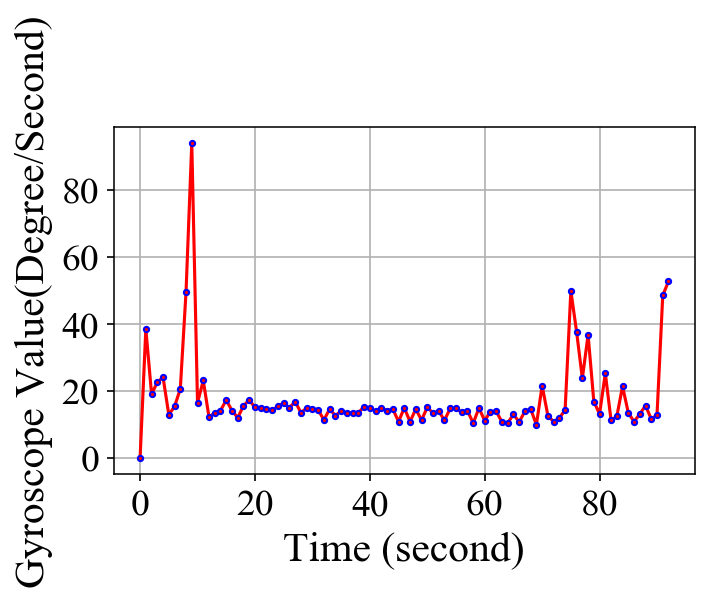}}%
\\
\subcaptionbox{Accelerometer data}{\includegraphics[width=0.23\textwidth]{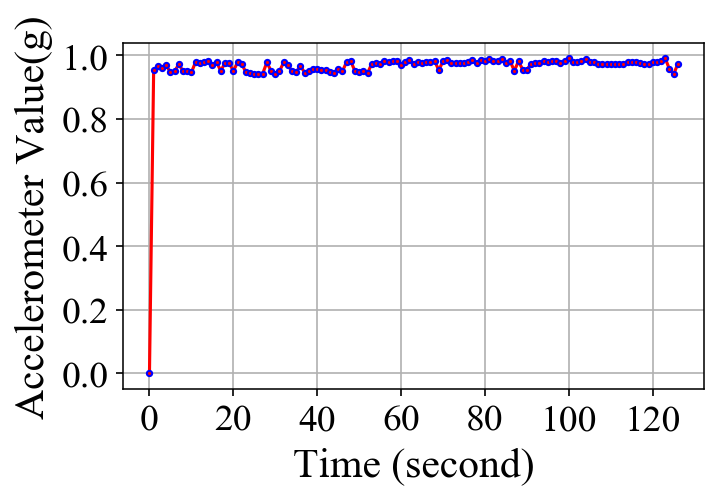}}%
\hfill % <-- Line break
\subcaptionbox{Gyroscope data}{\includegraphics[width=0.23\textwidth]{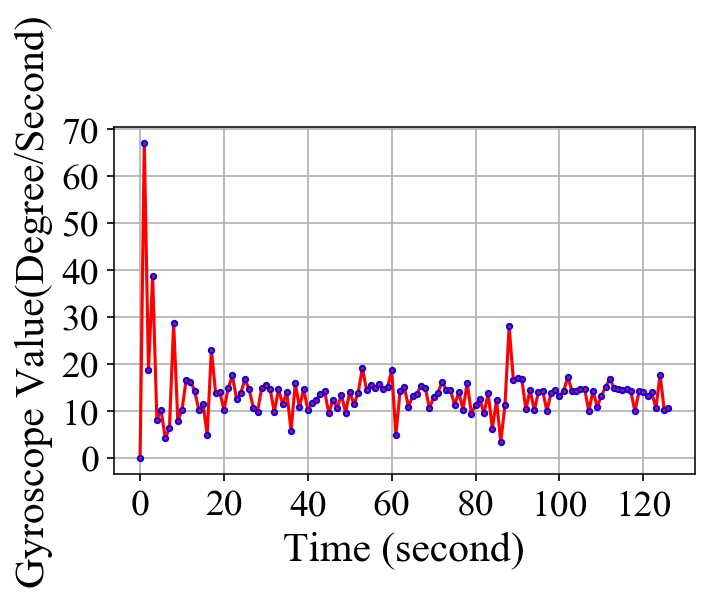}}%
\\
\subcaptionbox{Accelerometer data}{\includegraphics[width=0.23\textwidth]{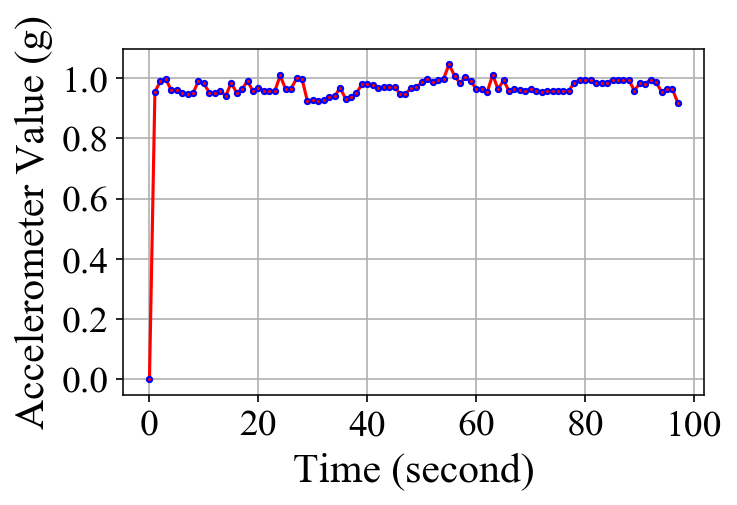}}%
\hfill % <-- Line break
\subcaptionbox{Gyroscope data}{\includegraphics[width=0.23\textwidth]{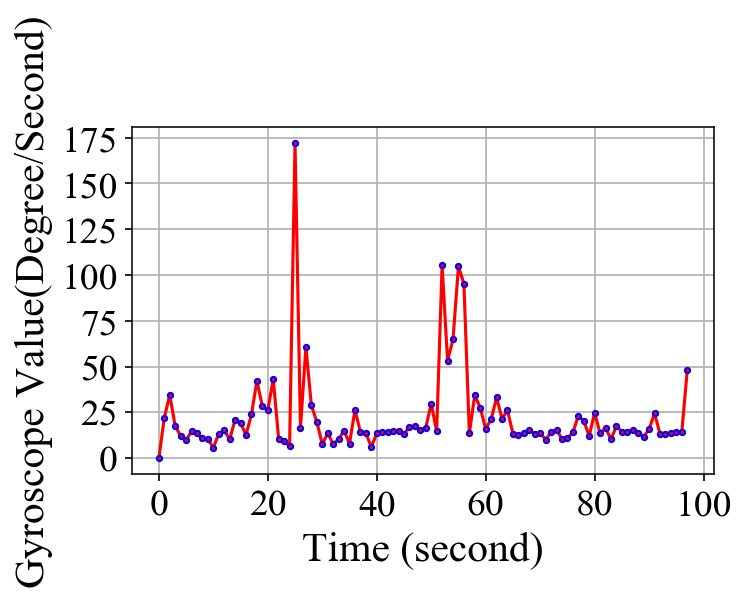}}%
\\
\subcaptionbox{Accelerometer data}{\includegraphics[width=0.23\textwidth]{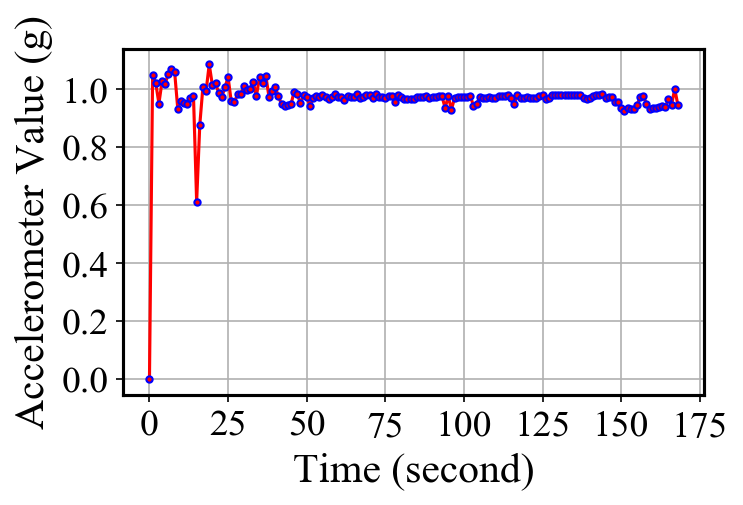}}%
\hfill % <-- Line break
\subcaptionbox{Gyroscope data}{\includegraphics[width=0.23\textwidth]{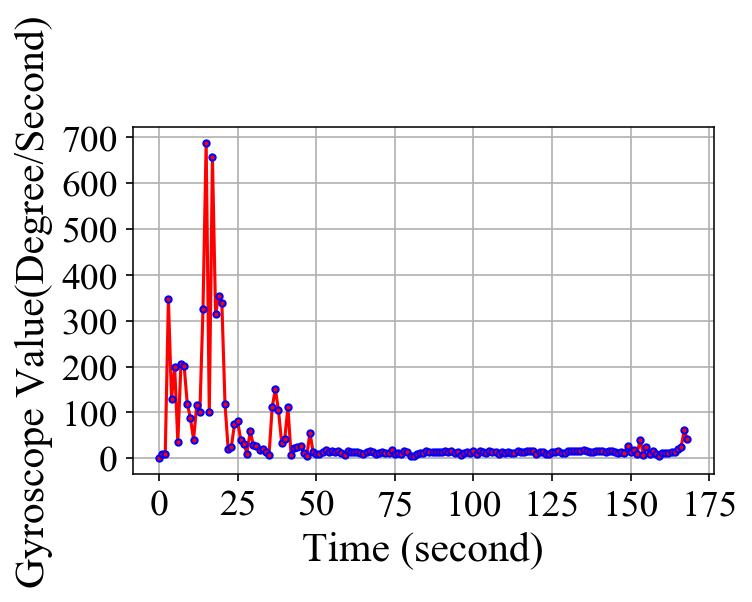}}%
\\
\subcaptionbox{Accelerometer data}{\includegraphics[width=0.23\textwidth]{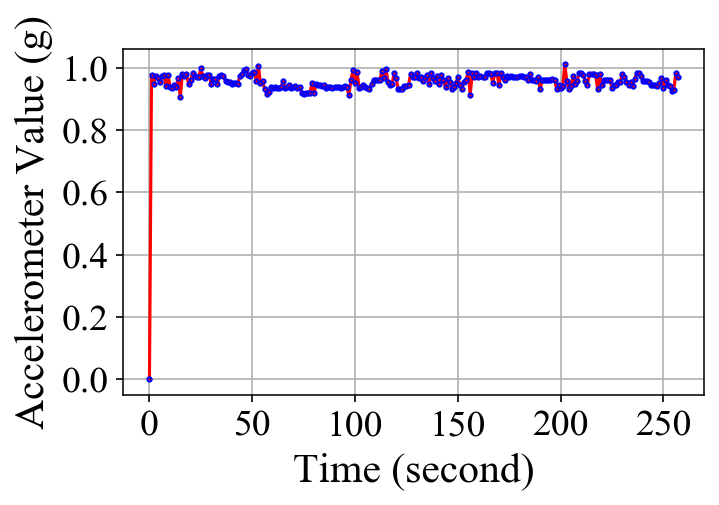}}%
\hfill % <-- Line break
\subcaptionbox{Gyroscope data}{\includegraphics[width=0.23\textwidth]{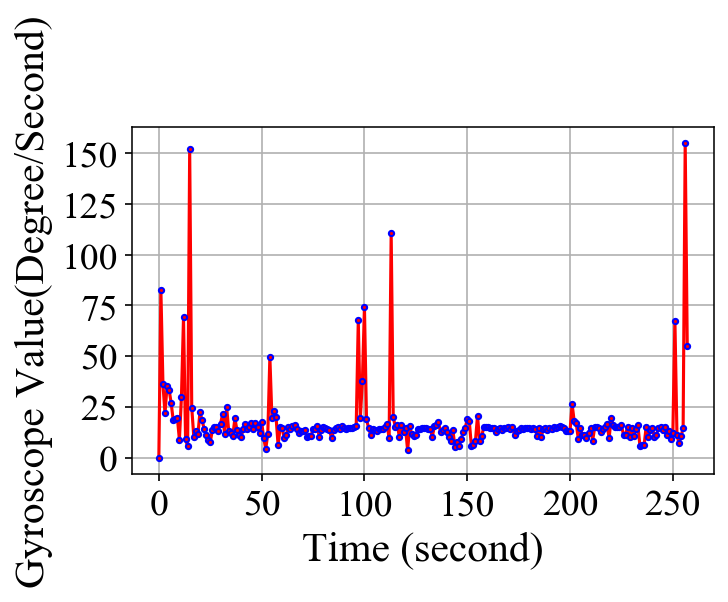}}%
\caption{Sensor data of every second of a random user. (a) and (b) are during showing neutral video,(c) and (d) are during showing Sad video;(e) and (f) are during showing video with disgust;(g) and (h) are during showing happy video,(i) and (j) are during showing video of surprise.}
\label{fig:sample}
\end{figure}

Our hypothesis was - \textit{There will be changes of head movement patterns in different emotional states}. Figure \ref{fig:sample} shows the results of head movement data (both components translational, rotational) in three different representative states. Figure \ref{fig:sample}: e,f depicts the disgust state data, whereas Figure \ref{fig:sample}: c,d shows data samples of sadness induced period. Figure \ref{fig:sample}: i,j is the depiction of data collected during another time-frame with surprise emotional state. All these data match with our hypothesis.

\subsection{Results for Validation}
We go one step further to train and test validity of machine learning models as mentioned in Section 3. This would enable us to extract meaningful information, features from head movement data pattern.
The best suited algorithm chosen over several machine learning algorithms demonstrate that our proposed technique achieves reasonable high training accuracy. However, our proposed technique demand significantly less memory and less computation time compared to the other algorithms making it suitable for pragmatic application. We have set \textit{success rate} as the acceptance criteria to compare among algorithms and take the one giving the best result.

As we mentioned in the previous section, we train 35 machine learning models for trait prediction and 1 model for emotion prediction. Here we describe the training accuracy of each of the 35 training models for trait prediction in Table: \ref{tab:table2}. The model for emotion prediction gives accuracy of 100{\%} under AdaBoostM1 Classifier with arguments [-P, 99, -I, 20, -Q, -S, 1, -W, weka.classifiers.trees.J48, --, -O, -S, -M, 1, -C, 0.9585694907535297]. 

The weighted average of performance related values for the model is given in table {\ref{tab:table1}}
\begin{table}[!t]
    \centering
    \begin{tabular}{|c|c|c|c|c|} \hline 
        \textbf{TP Rate} & \textbf{FP Rate} &
        \textbf{Precision} &
        \textbf{Recall} &
        \textbf{F-measure} 
        \\ \hline
        100\% & 0\% & 100\% & 100\% &100\%
        \\ \hline
        
\end{tabular}
    \caption{Performance related value of model for emotion prediction} 
    \label{tab:table1}
\end{table}

\begin{table}[!t]
    \centering
    \begin{tabular}{|l|c|} \hline 
        \textbf{Predicted Trait - Emotion  } & \textbf{Accuracy(\%)}\\ \hline
        Smoker - Happy & 97.5 \\ \hline
        Smoker - Neutral & 97.5 \\ \hline
        Smoker - Surprise & 97.5\\ \hline
        Smoker - Sad & 97.5 \\ \hline
        Smoker - Disgust & 97.5 \\ \hline
        Practitioner - Happy & 80 \\ \hline
        Practitioner - Neutral & 80 \\ \hline
        Practitioner - Surprise & 62.5 \\ \hline
        Practitioner - Sad & 90 \\ \hline
        Practitioner - Disgust & 92.5 \\ \hline
        Fast food intake - Happy & 87.2 \\ \hline
        Fast food intake - Neutral & 92.3 \\ \hline
        Fast food intake - Surprise & 66.7 \\ \hline
        Fast food intake - Sad & 69.23\\ \hline
        Fast food intake - Disgust & 71.8 \\ \hline
        Fat intake - Happy & 82.5 \\ \hline
        Fat intake - Neutral & 62.5 \\ \hline
        Fat intake - Surprise & 87.5 \\ \hline
        Fat intake - Sad & 100 \\ \hline
        Fat intake - Disgust & 55 \\ \hline
        Sugar intake - Happy & 95 \\ \hline
        Sugar intake - Neutral & 97.5 \\ \hline
        Sugar intake - Surprise & 87.5 \\ \hline
        Sugar intake - Sad & 97.5 \\ \hline
        Sugar intake - Disgust & 97.5 \\ \hline
        Heart disease in family - Happy & 100  \\ \hline
        Heart disease in family - Neutral & 100 \\ \hline
        Heart disease in family - Surprise & 92.5 \\ \hline
      Heart disease in family - Sad & 85 \\ \hline
        Heart disease in family - Disgust & 100 \\ \hline
        Diabetes in family - Happy  & 67.5 \\ \hline
        Diabetes in family - Neutral & 67.5 \\ \hline
        Diabetes in family - Surprise & 80 \\ \hline
        Diabetes in family - Sad & 67.5 \\ \hline
     Diabetes in family - Disgust & 72.5 \\ \hline
    \end{tabular}
    \caption{Accuracy of training models} 
    \label{tab:table2}
\end{table}
%will update it as all results come.-Purabi

\begin{table*}[!t]
    \centering
    \begin{tabular}{|c|c|c|c|c|c|c|}
        \hline 
           \textbf{Predicted trait-Emotion} & \textbf{Best classifier}& \textbf{TP rate(\%)}& \textbf{FP rate(\%)} & \textbf{Precision(\%)} & \textbf{Recall(\%)} & \textbf{F-measure(\%)} \\ \hline
           Smoker-Happy & RandomForest & 97.5 & 11.8 & 97.6 & 97.5 & 97.4 \\ \hline
           Smoker-Neutral & SMO & 97.5 & 11.8 & 97.6 & 97.5 & 97.4 \\ \hline 
           Smoker-Surprise & Logistic & 97.5 & 11.8 & 97.6 & 97.5 & 97.4 \\ \hline 
           Smoker-Sad & AdaBoostM1 & 97.5 & 11.8 & 97.6 & 97.5 & 97.4 \\ \hline 
           Smoker-Disgust & RandomForest & 97.5 & 11.8 & 97.6 & 97.5 & 97.4 \\ \hline
           Practitioner-Surprise & AdaBoostM1 & 62.5 & 56.3 & 76.9 &62.5 & 50.4 \\ \hline
           Practitioner-Neutral & RandomTree & 80 & 25.8 & 80.5 & 80.0 & 79.3 \\ \hline
           Practitioner-Happy & JRip & 80 & 21.7 & 80.0 & 80.0 & 58.3 \\ \hline
           Practitioner-Sad & Random Tree & 90 & 12.9 & 90.2 & 90.0 & 89.9 \\ \hline
           Practitioner-Disgust & Vote & 92.5 & 5.0 & 93.7 & 92.5 & 92.6 \\ \hline
           Fastfood intake-Surprise & LWL & 66.7 & 32.9 & 76.6 & 66.7 & 62.5 \\ \hline
           Fastfood intake-Happy & LWL & 87.2 & 9.2 & 89.0 & 87.2 & 86.7 \\\hline
           Fastfood intake-Neutral & RandomTree & 92.3 & 8.1 & 93.3 & 92.3 & 92.0 \\\hline
           Fastfood intake-Sad & LWL & 69.2 & 28.7 & 73 & 69.2 & 67.5 \\\hline
           Fastfood intake-Disgust & OneR & 71.8 & 16.6 & 74.4 & 71.8 & 72.2 \\\hline
           Fat intake-Happy & AdaBoostM1 & 82.5 & 15.3 & 85.0 & 82.5 & 82.4 \\ \hline
           Fat intake-Surprise & LWL & 87.5 & 12.2 & 87.0 & 87.5 & 87.5 \\ \hline
           Fat intake-Sad & REP Tree & 100 & 0 & 100 & 100 & 100 \\ \hline
           
           Fat intake-Neutral & SMO & 62.5 & 43.8 & 65.5 & 62.5 & 58 \\ \hline
           Fat intake-Disgust & RandomCommittee & 55 & 55 & NAN & 55 & NAN \\ \hline
           High Sugar intake-Surprise & J48 & 87.5 & 12.8 & 87.8 & 87.5 & 87.6 \\ \hline
           High Sugar intake-Neutral & REPTree & 97.5 & 4.2 & 97.6 & 97.5 & 97.5 \\ \hline
           High Sugar intake-Happy & RandomForest & 95 & 5.7 & 95 & 95 & 95 \\ \hline
           High Sugar intake-Sad & IBK & 97.5 & 4.2 & 97.6 & 97.5 & 97.5 \\ \hline
           High Sugar intake-Disgust & OneR & 97.5 & 4.2 & 97.6 & 97.5 & 97.5 \\ \hline
           Heart Disease-Meditate & OneR & 100 & 0 & 100 & 100 & 100 \\ \hline
           Heart Disease-Happy & LWL & 100 & 0 & 100 & 100 & 100 \\ \hline
           Heart Disease-Surprise & LWL & 55 & 45 & 55 & 55 & 55 \\ \hline
           Heart Disease-Disgust & AttributeSelectedClassifier & 100 & 0 & 100 & 100 & 100 \\
            \hline
           Heart Disease-Sad & IBK & 100 & 0 & 100 & 100 & 100 \\\hline
           Diabetes-Happy & OneR & 67.5 & 67.5 & NAN & 67.5 & NAN \\ \hline
           Diabetes-Disgust & OneR & 72.5 & 49.1 & 71.3 & 72.5 & 69 \\ \hline
           Diabetes-Neutral & MultilayerPerception & 67.5 & 67.5 & NAN & 67.5 & NAN \\ \hline
           Diabetes-Sad & LWL & 67.5 & 67.5 & NAN & 67.5 & NAN \\ \hline
           Diabetes-Surprise & OneR & 80 & 25.6 & 80 & 80 & 80 \\ \hline
        \end{tabular}
    \caption{Classifier and Performance related value(weighted average) of training models} 
    \label{tab:table3}
\end{table*}

In table \ref{tab:table3} we provide classifiers and performance related values of the models that we used to get the accuracy given in table \ref{tab:table2}.
%barchart's new code.
\begin{figure*}[!t]
    \centering
     \begin{subfigure}{0.40\textwidth}
         \centering
         \includegraphics[width=\textwidth]{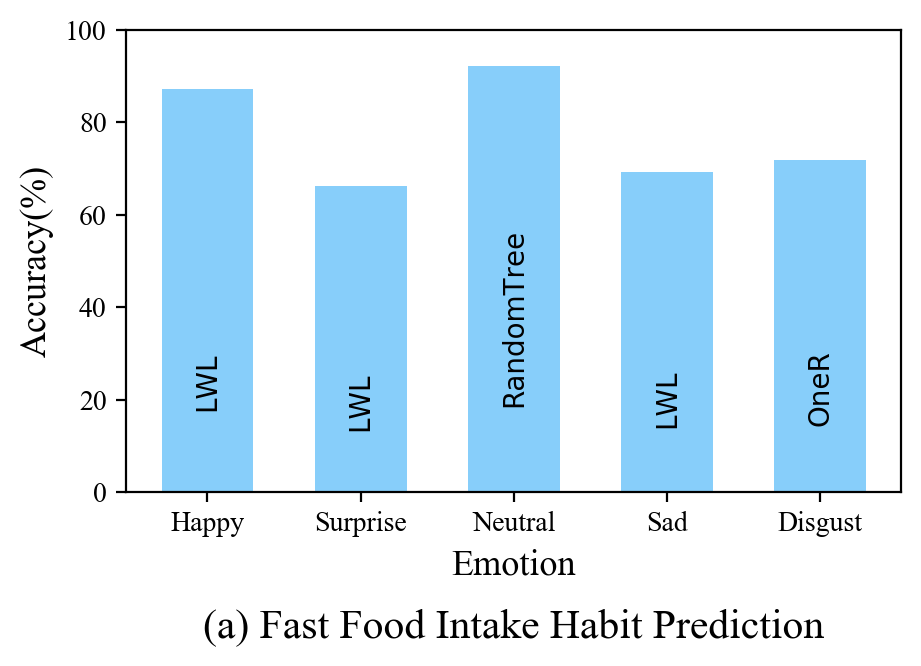}
         \caption{}
         \label{Happy1}
     \end{subfigure}
     \hfill
     \begin{subfigure}{0.40\textwidth}
     \centering
     \includegraphics[width=\textwidth]{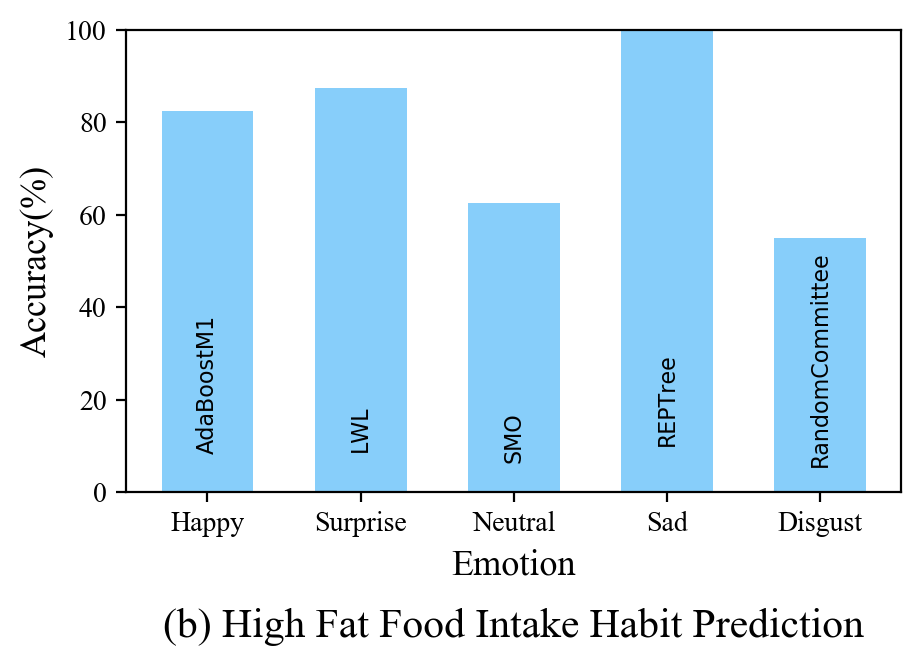}
     \end{subfigure}
          \begin{subfigure}{0.40\textwidth}
         \centering
         \includegraphics[width=\textwidth]{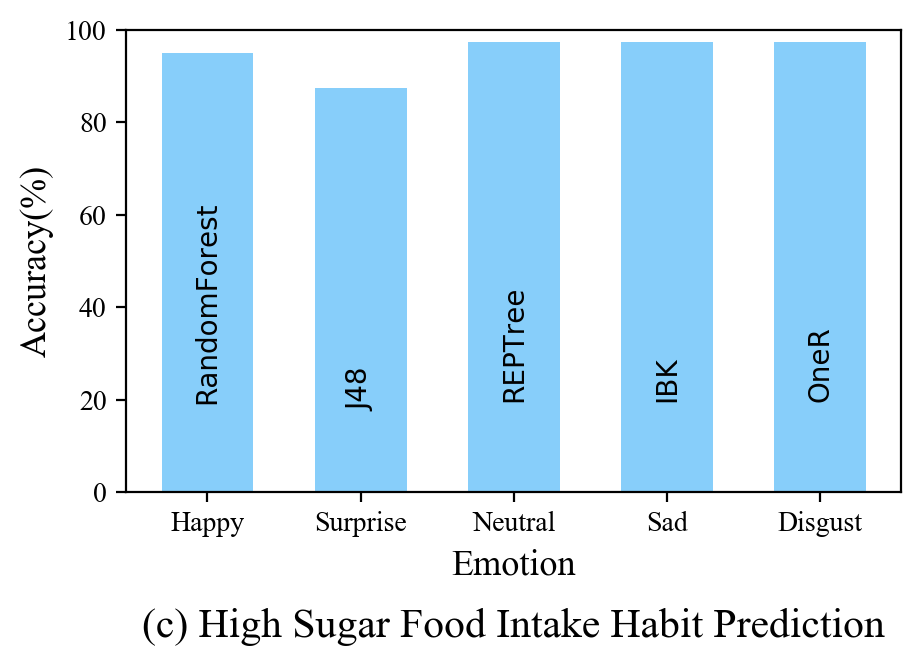}
         \caption{}
         \label{Happy2}
     \end{subfigure}
     \hfill
     \begin{subfigure}{0.40\textwidth}
     \centering
     \includegraphics[width=\textwidth]{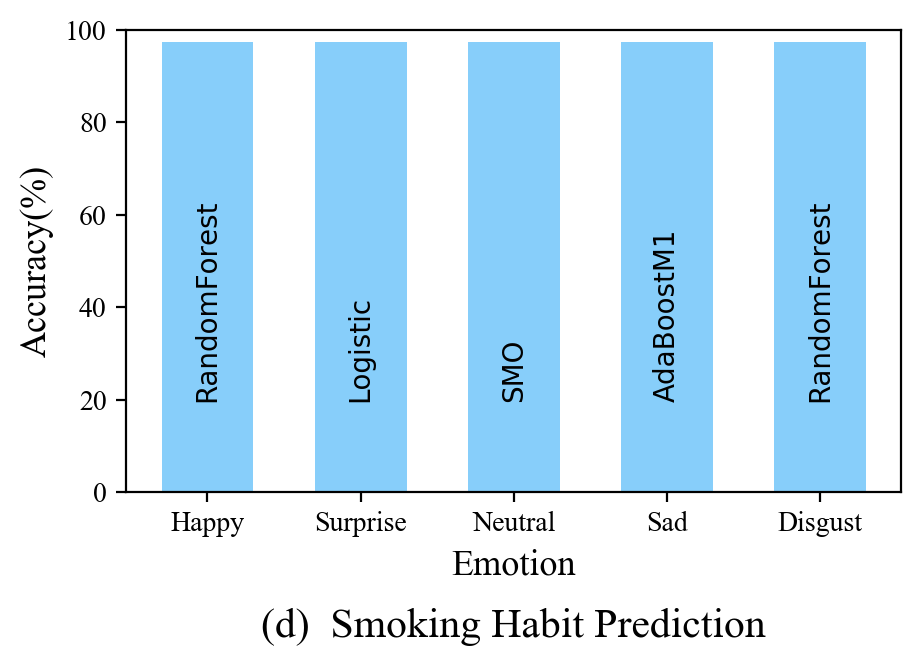}
     \end{subfigure}
     \begin{subfigure}{0.40\textwidth}
     \centering
     \includegraphics[width=\textwidth]{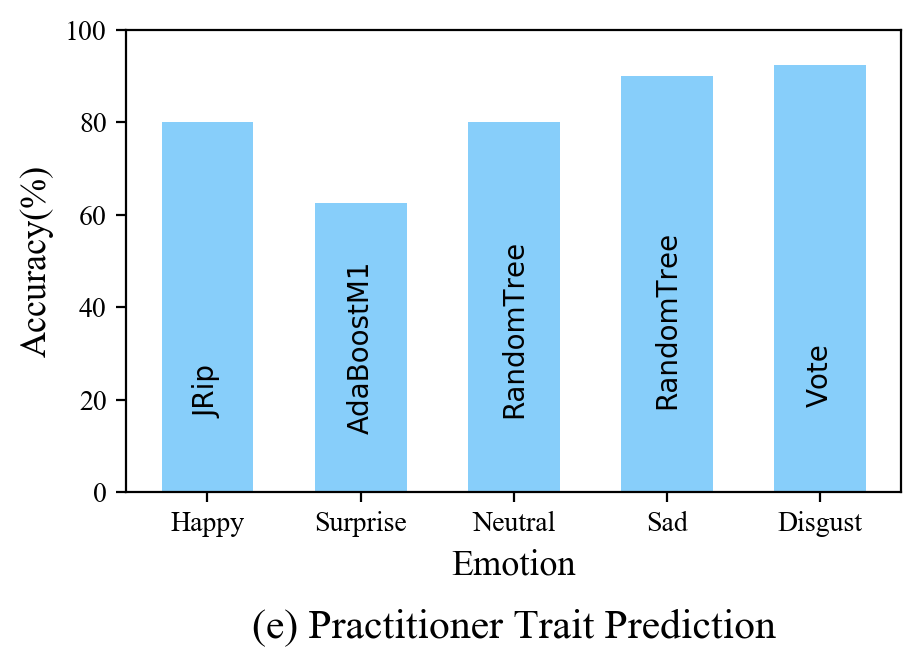}
     \end{subfigure}
     
\caption{Accuracy and best classifier for different emotion in identifying traits}

\label{fig:accu}

\end{figure*}

In Figure \ref{fig:accu} we provide the accuracy of each traits in each video and the classifier for it in bar chart format. The algorithm is written in the bar providing the figure-mentioned accuracy. In Figure \ref{fig:accu}:a the accuracy of fast food intake prediction is accumulated based on data taken at different emotional states. Figure \ref{fig:accu}:b for high fat food taking prediction. The prediction models for Smoking habit, fast food taking,high sugar taking habits,practicing religion habit are also shown in Figure \ref{fig:accu}. A more details are listed in table \ref{tab:table4}.

%mirajul(revised)
%purabi
\subsection{Hardware Requirement in Training Phase}
Through an evaluation with 46 users, our system shows reasonably high accuracy. To perform the identification task, the application takes only 12 minutes on an average which accumulates the time of showing 5 videos each having a duration of 1.5 to 3 minutes approximately. The application uses 3-4\% of CPU with 100-105 MB memory on the laptop during the small period of video showing and answering survey questions from participants.The accelerometer and gyroscope data are sampled in one second interval at the same time from our android application without causing any trouble to the participants while watching video.

%mirajul(revised)
%purabi

\begin{table*}[!t]
    \centering
    \begin{tabular}{|l|c|}
        \hline 
           \textbf{Predicting Questions}  & \textbf{Best Accuracy(\%)} \\ \hline
           Are you a smoker?    & 97.5  \\
          \hline
          Do you pray regularly? & 92.5       \\
          \hline
          How many fast food meals in a weak?    &  92.3\\
          \hline
          Do you take high fat food? & 100  \\
          \hline
          Do you take high sugar food? & 97.5 \\ 
          \hline                                  
          Does any body in your family have heart disease?  &  100\\
          \hline
          Does any body in your family have Diabetes? & 80\\
          \hline
        \end{tabular}
    \caption{Survey questions and summary of the prediction evaluation} 
    \label{tab:table4}
\end{table*}

\subsection{Accuracy of Different Predictions}
We have measured the prediction accuracy of some traits after training the collected data from the participants. From Table \ref{tab:table4}, we can see the accuracy of the predictions. The maximum accuracy of all emotional states has been reported in Table \ref{tab:table2} for each trait.

\subsection{Threats to Validity}
Our research work has been conducted in a controlled laboratory environment. There is a chance that the same set of results, correctness may not be on the line when tested in noisy environment. Moreover, the respondents to our survey system have no disabilities or any diseases related to abnormal head shaking or movement. Experimentation with these specialized participants will probably not bring the same set of conclusions.

\section{Conclusion and Future Work}
This paper contributes in the field of human trait and emotional state identification by means of head movement analyses using different sensors and machine learning techniques. This research does not only find different head movement patterns in different induced emotional states, but also finds correlations between physical traits and head movement patterns.  

In future, we will increase the number of participants to enlarge our data set to enhance the accuracy of our identification. Besides, to scale it up more, we will generate synthetic data based on our real collected data to make our training set larger. Additionally, even though we have explored 14 human traits, there are many more human traits that might have correlation with head movement. We plan to explore these in future.In these 14 traits we could only make balanced data set for 7 traits. We will try to make remaining trait's(religious belief,sleep problem,alcoholic etc.) data set balanced so that we can analyse these to find out if there is correlation between these traits and head movement. 

There exist other avenues to work with the head movement sensing system with people having specific head movement patterns due to illness or habits. The patterns may lead to deeper understanding about the reasons lying behind these patterns. We plan to work on investigating the patterns in future. Finally, we also plan to explore head movement of visually-impaired people to explore their traits and emotional states to progress this research further. 

\section{Acknowledgment}
This work has been conducted in Bangladesh University of Engineering and Technology (BUET), Dhaka, Bangladesh. Besides, the authors are thankful to Dr. Fahim Kawsar, Nokia Bell Lab for providing eSense devices that were used in this study.

\bibliographystyle{ACM-Reference-Format}
\bibliography{sample-base}

\end{document}